# Expression driven Trignometric based Procedural Animation of Quadrupeds

Zeeshan Bhatti, Asadullah Shah, Mustafa Karabasi and Waheed Mahesar
Khulliyyah of Information and Communication Technology
International Islamic University Malaysia, Kuala Lumpur
e-mail: zeeshan.bhatti@live.iium.edu.my, asadullah@iium.edu.my, mostafa.karabasi@live.iium.edu.my,
abdul.waheed@live.iium.edu.my

*Abstract*— This research paper addresses the problem of generating involuntary and precise animation of quadrupeds with automatic rigging system of various character types. The technique proposed through this research is based on a two tier animation control curve with base simulation being driven through dynamic mathematical model using procedural algorithm and the top layer with a custom user controlled animation provided with intuitive Graphical User Interface (GUI). The character rig is based on forward and inverse kinematics driven through trigonometric based motion equations. The User is provided with various manipulators and attributes to control and handle the locomotion gaits of the characters and choose between various types of simulated motions from walking, running, trotting, ambling and galloping with complete custom controls to easily extend the base simulation as per requirements.

*Keywords; Quadruped Animation ; simulation; procedural animation; trigonometric expressions;*

## I. INTRODUCTION

Animation is a very frequently used term in computer graphics and multimedia, generally considered to be a synonym to motion. However the current form of computer animation covers a wide range of motions concerning changes that have a dynamic visual effects including shape, color, transparency, structure, texture of an object, camera position, orientation and focus along with position, orientation and dynamics of objects on the timescale [1]. Animation thus, deals with the process of simulating real and dynamic behavior of objects and events using roles of computer graphics, multimedia and the physics of particle dynamics [1][2].

The use of animation has advanced immensely in this last decade in the field of Entertainment, Games, Movies, Advertisement, Scientific and Architectural Visualizations. Among various types, the character animation has always been the most significant and demanding of all. It is used to breathe life into virtual character and to make them act like real actors. The procedure of bringing a virtual character to life using various animation techniques is lengthy and monotonous process.  There are numerous commercial animation tools available in the market with vast set of tools implementing diverse kinds of skeletal rigs for animating virtual characters. The tool sets available in such type of software's are very efficient with vide variety but usually they seem to fail in satisfying the computer animator or riggers need [3]. So a character animator normally ends up building a custom skeletal rig for the ease of animation [3][4]. The process is also known as Character Rigging. We have used MAYA as the basic development and simulation tool with motion equations implemented as Maya Embedded Language (MEL) code written as expressions that gets executed at every frame controlling and driving the various body part of the character rig.

## II. LITERATURE REVIEW

Quadruped motion has always been an integral part of character animation and simulation. Various techniques and approaches have been used to understand the realistic motion gaits and parameters of quadruped animals. In 3D computer graphics area, the Inverse Kinematics (IK) based approach for leg transformation is used widely as discussed by Kokkevis [5] and Torkos [6], where the placement of the foot is decided first and then physical based models drives the body. Whereas an advance technique of producing real-time interaction based simulation of quadrupeds, is through controller-based approach as discussed by S. A. Marsland [7] and S. Coros [8].

3D forward dynamic simulations of quadruped gaits are introduced by Raibert and Hodgins [2], who developed control strategies for trotting, bounding, and galloping gaits for a robot quadruped with a rigid body and extensible legs. Similarly the game of Spore by Hecker, C. [9] develops methods for generating procedural animation for arbitrary legged creatures, including locomotion patterns. However Ljiljana Skrba in their paper "Quadruped Animation" discusses various ways for achieving realistic quadruped motion, through "video-based acquisition, physics based models, inverse kinematics," or some combination of the above [10]. Whereas the methodology proposed by Skrba is also dependent on video based input and lacks easy to use system with custom user interaction. Conversely Wampler and Popović [11] develop a two-level optimization procedure for physics-based trajectories of periodic legged locomotion and use it to explore connections between form and function. Whereas Kwon and Shin [12] have analyzed the center-of-mass trajectory of human walking and running motions to segment unlabeled motion sequences into motion half-cycles. On the other hand Stelian Coros [8] in his paper discuss physics based simulation centered on integrated set of gaits and skills covering a wide range of motion repertoire of a dog.

## III. PROBLEM DEFINATION

Animation in 3D applications usually happens in two primary ways, through Key-frame animation and Motion Capture. Key-framing is the oldest and traditional technique which is immensely popular and still being widely used in the animation industry. The process of animating a character using key-frame technique is very long and tedious. In this the animator has to manually move and reposition each body part through certain controllers and manipulators, positioning them with respect to time and keying their attributes. This process is lengthy and monotonous. The other technique used for animation is through Motion Capture. There are various downsides to animating through motion capture (mocap). First the cost of mocap technology is huge, which can be around thousands of US dollars, thus making it very difficult for an ordinary production studio or freelance 3D animators to implement. Secondly, the animation artist then has to learn the tools and techniques of importing the mocap data and incorporating it onto the 3D character rig. Thirdly, problem with motion capture is that the mocap animation of quadrupeds is very difficult and practically impossible to get in various situations. The major downside of using motion capture based animation is the obtained result from sensory data is far from perfection and eventually the animator has to perform various clean-up operations to remove jitter and noise from the animation.

The animation of articulated characters such as Tigers, Lions or Horses, possess a major challenge as each type of character have vast range of gaits and motion types with joints having multiple degrees of freedom. Key-framing being the traditional technique provides support only at a low level- the animator tediously creates various control rigs to control and manipulate joint angles or coordinates for each bone type and its degree of freedom.

### A. Research Aim

This research aims to work on developing a system which will make animating quadrupeds a very comfortably and automated process through the use of procedural animation. In order to simplify and automate the process of character animation, the use of higher level technique called procedural animation is implemented, where trigonometric based mathematical expressions are used to achieve automated movements for animating each body part of quadruped characters, creating realistic looking animation easily and quickly. By using this procedural technique the animator will have all the basic animation of a character at a single click and then after that, the animator can customize the animation according to his story board and requirement, through automatically created control rig with custom manipulators as proposed in [3].

## IV. QUADRUPED RIG MODEL

Initially an abstraction of the quadruped skeletal structure is proposed with single joint chain representing each of the four legs of the character facilitating each foot placement calculations independently as proposed by Coros at. el. [8]. We then further enhanced this proposed model and added few additional joints for various body parts. For front legs a single additional Carpus joint is introduced for knee bending. At hind legs two joints are introduced for Pattella and Calcaneus joint locations for simulating more natural leg behavior. A flexible abstract spine multi-joint model has been developed which helps achieve more natural motion during fast gaits such as the gallop. The flexible spine connects front and back leg frames, which together form a dual leg frame model, with each leg frame making independent decisions regarding the footstep placement, height control, and pitch control [8]. In tail and neck we have also added few additional joint to produce accurate rotational arcs during animation. Once the abstract model based mathematical algorithm has been achieved, then more specific and standard model is derived and equations are modified to produce simulation for a standard skeletal model as discussed in [3] [14]. Then the actual 3D model is created representing the physical skeletal structure of a Horse. This skeletal joint chain is then duplicated and constrained to the original base skeleton, thus we now have two identical skeletal structures providing a dual layer based approach. The base skeletal structure is used for rigging whereas custom controls are provided for user manipulation according to the animators need. The second skeleton is used for procedural simulation and is driven through the trigonometric equations. This system of dual layer quadruped model is illustrated in figure 1.

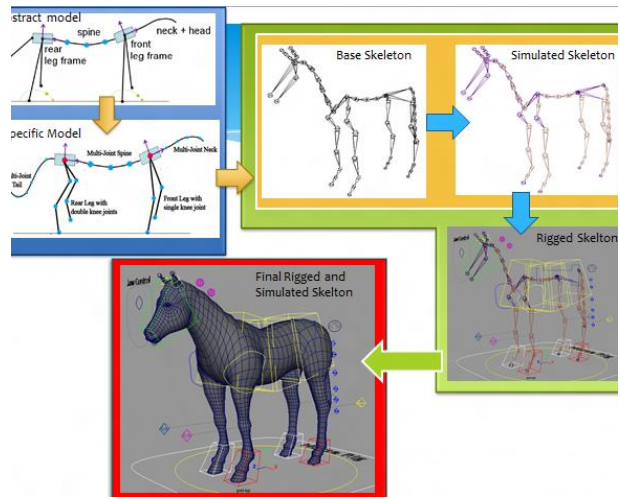

Figure 1: Quadruped dual layer rigged character model

## V. MATHEMATICAL MODELING

A mathematical model has been developed in which the control representation is compact, expressive, and provides robust cyclic motion. Figure 2 provides the overview of the mathematical model for the procedural animation of quadrupeds. The overall control loop is driven using Motion Frequency ($M_{freq}$) parameter which provides torques to a forward dynamics simulator at every time step. The motion parameters of each body part is calculated and generated individually. The spine motion is calculated using the sine function as shown in (1) already discussed in [15].

$$\text{SpineSwing } S_{sw} = \sin(tf * M_{freq} * C_{error}) * M_{osc} \quad (1)$$

The current time in frame (*tf*) is obtained from the timeline and C$_{error}$ is a constant used for counter gait error for optimization. The oscillation of each joint during motion is controlled by M$_{osc}$, providing the flexibility and bounce in the motion. The head and tail motion curves are also obtained by using the trigonometric sine function with motion frequency and optimizing with joint error (J$_{error}$) constant as given in (2).

$$\sin(tf * M_{freq} * J_{error}) \quad (2)$$

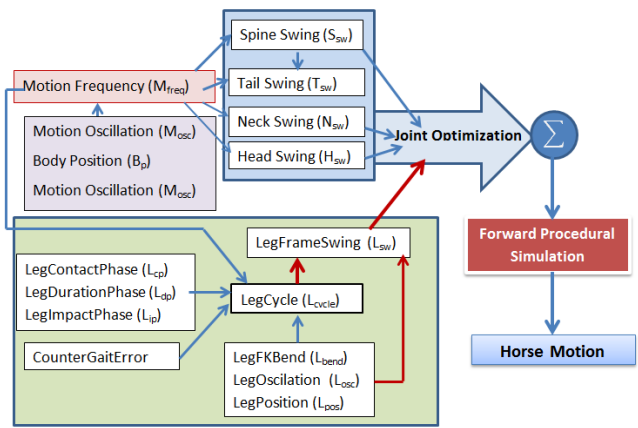

Figure 2: Overview of controller based mathematical model.

The joint chain of each leg is initially constrained and attached to Inverse Kinematics (IK) based controller for achieving realistic motion. Each leg motion is calculated individually with leg impact phase (L$_{ip}$) and leg contact Phase (L$_{cp}$) controlling the gait position while leg duration phase(L$_{dp}$) determines the time the leg is in motion. The IK based translation of the each leg is obtained using (3). The FPS is a constant variable used for the frame per second count, and L$_{cycle}$ is calculated from number of cyclic leg motion of a horse per second.

$$\text{LegSwing } (L_{swa}) = \frac{P_{id} - 2 * \pi * FPS + L_{cycle} - L_{rp}}{L_{cycle}} \quad (3)$$

The sternum joints of the horse Skelton are moved in the corresponding 'z' direction using the cosine trigonometric function with counter gait error (C$_{error}$) and Joint error optimization parameters multiplied with the desired leg oscillation as given in (4).

$$\text{LegSternum.tZ} = \cos(tf * M_{freq} - (\pi/4)) * L_{osc} * C_{error} + J_{error} \quad (4)$$

*A. Leg Motion Cycle*

The motion cycles considered here are cyclic walk or amble cycles, where the character is walking on the ground beneath with a constant velocity. The same applies to motion with turning, where a character is turning around while walking [13]. There can be any number of legs and the movements of the different legs are analyzed completely independently from each other. For each leg a heuristic is used to find a stance time (The point in time in the motion cycle when the leg is standing most firmly and neutrally on the ground) as shown in figure 3.

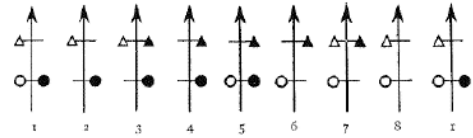

Figure 3: Phases of the walk for a horse

Time is defined cyclically for the motion cycles, with a value of 0.0 denoting the start of the cycle and the end of cycle is represented by 1.0, after which the cycle starts over again. In addition to the overall motion cycle, each leg has its own leg cycle as shown in figure 4.The start and end of these leg cycles are marked by the stance times of the respective legs, the start times of the leg cycles are not arbitrary at all [13].

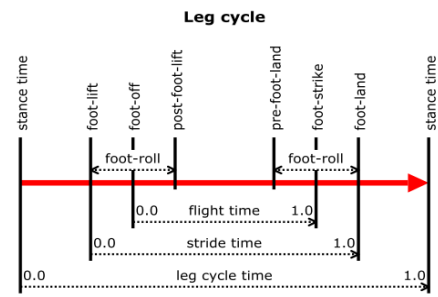

*Figure 4*: Leg Motion Cycle from stance to motion to stance.[13]

## VI. RESULTS

The use of simple trigonometric functions and algebra provided us with a basic set of equations that can be easily used for generating high end quadruped motion. Figure 5 shows the skeletal model of a horse character built on inverse kinematics, with elementary set of rigging controls. We used this skeletal joint hierarchy to create the dual layered based animation setup. The first layer of joints chain is controlled by the trigonometric expressions and the second layer provides user controls for custom manipulation of the animated character. Using this model we can produce several animation types for quadruped characters including walking, trotting, ambling and racking. The gaits transition was achieved using the Leg Impact Phase (Lip) and duration (Ldp) of contact with the ground.

Inverse Kinematics (IK) system is used to control joint chains of the spine, tail and neck individually. Each IK is driven through expressions that define the oscillation motion curves according to the gaits. For leg joint each IK were used with expressions controlling the translating attribute. The joint of the foot are separately controlled using simple forward kinematics. An easy to use graphical user interface was designed in MAYA that provided the basic control over the simulation.

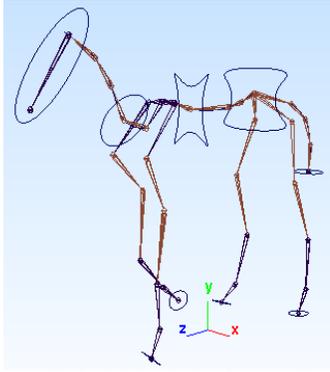

Figure 5: Skeletal joint chain of Horse Character

Figure 6(a) shows the parameters used to control the motion of the body. The 'Speed' attribute determines the overall motion frequency of the characters. The 'High' attribute is the height of the head, neck and tail with respect to the spine and hip joint of the character during motion. The 'Bounce' attribute controls the overall bounce in the characters body during any type of motion.

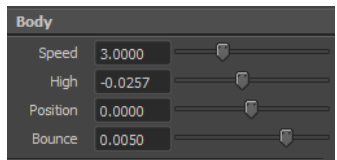

Figure 6(a): Main body parameters

The figure 6(b) shows the head and tail attributes. The head motion is controlled by three parameters; 'Head High' controls the height of head with 'Head Pos' determines the position of head with respect to its orientation from neck. The 'Head Oscillation' controls the number of oscillations of head joints during any type of motion. Similarly the tail swing attribute controls the swing frequency of the tail during motion. All of these attributes can be changed dynamically during runtime to change the behavior of the quadruped motion.

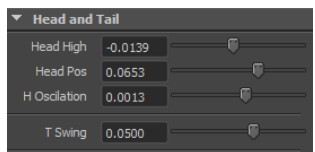

Figure 6(b): Head and Tail parameters

The parameters for the leg motion are shown in figure 6 (c to f). Each legs is controlled individually by set of same parameters. All for legs have different values for different type of motion gaits. The parameters shown in figure 6(a-f) are for walk motion gaits. The leg impact phase and impact duration controls the main parameter between various different gaits. These attribute determines the time it takes for each leg to touch the ground and the duration for which each leg remains in contact with the ground.

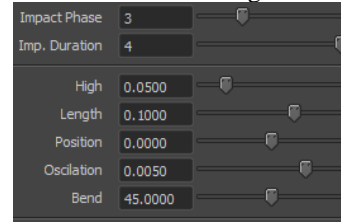

Figure 6 (c): Front right leg parameters

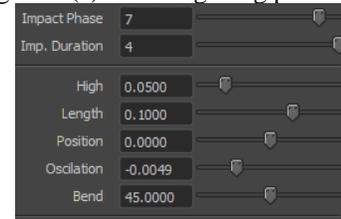

Figure 6(d): Front left leg parameters

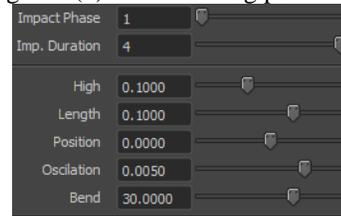

Figure 6(e): back right leg parameters

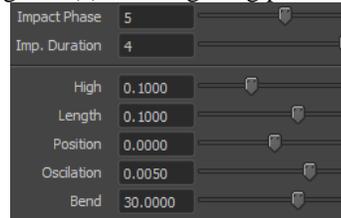

Figure 6(f): back left leg parameters

The animation is created by modifying these attributes according to the characteristic of each motion gait during animation. A preset library has been created to transition between various motion gaits. Figure 7 shows the procedurally generated animation of horse during Walk and figure 8 shows the various during Amble motion. For amble motion the motion frequency was set to 4, with impact duration of each leg set to 3. The impact phase of each leg was set to 1 for front right leg, 5 for front left leg, 7 for back right leg, and 3 for back left leg.

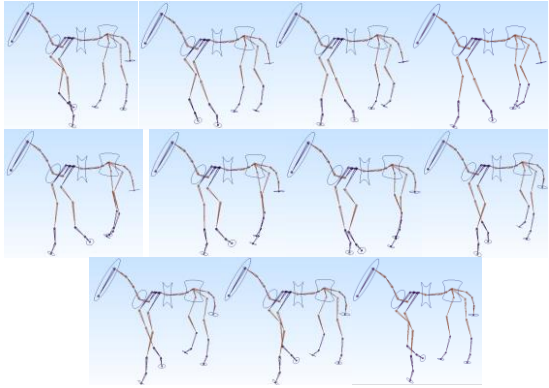

Figure 7: Phases of walk generated through expressions

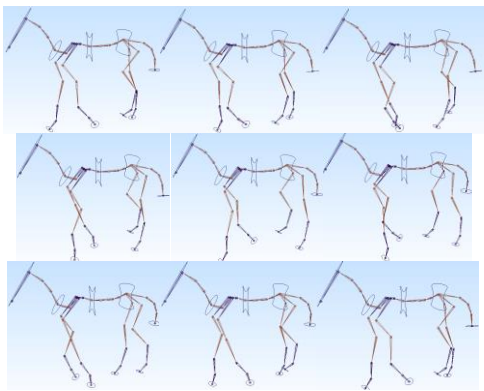

Figure 8: Amble motion of Horse

## VII. CONCLUSION AND FUTURE WORK

We have developed a new technique of dual layer based animation model, with base layer driven through procedural expressions using trigonometric functions, and the top layer is controlled manually by custom user controlled manipulators. Through mathematics exact animation of character gaits, locomotion behavior and various motion processing aspects are directly incorporated into the control algorithms. The motion is then controlled by an easy to use animator friendly GUI, with simplified attributes that internally effect the expression values driving the articulated character skeleton. Through this technique the process of animating quadrupeds is simplified and the base motion is generated easily. After the motion is achieved the animator then can easily grab the custom manipulators and modify the animation as per need and requirement. This approach will greatly enhance the quality and reduce the time consumed for creating animation from scratch.

This work needs to further improve the skeletal model, which has to be generalized to fit various other quadruped characters like Lion, Tiger, Dog, Zebra, etc. Various variables have to be inducted that could affect the behavior and pattern of quadruped motion cycle. The control mechanism for terrain elevation and change in motion direction is also covered in this project which further needs to be researched and implemented. A rigorous testing mechanism has to be implemented to ensure the generated motion is absolutely accurate and realistic. More motion gaits have to be analyzed and incorporated into the algorithm with advanced set of custom control manipulators to produce high quality animation.

## ACKNOWLEDGEMNT

This work is partially funded my Ministry of Higher Education Malaysia (MOHE) under Commonwealth Scholarship and Fellowship Program.